\documentstyle[epsfig,aps,12pt]{revtex}
\def\bge{\begin{equation}}
\def\ene{\end{equation}}
\def\bg{\begin{eqnarray}}
\def\en{\end{eqnarray}}
\def\nn{\nonumber}

\def\D0bar{\overline{D^0}}

\begin{document}
\tighten
\title{Bound Nucleon Form Factors, 
Quark-Hadron Duality, and Nuclear EMC Effect}

\author{
K. Tsushima$^1$\footnote{Talk presented by K.~Tsushima at the Joint
JLab-UGA Workshop
on {\it ``Modern Sub-Nuclear Physics and JLab Experiments"}, 
to honor the occasion of Dr. Joe Hamilton (Vanderbilt University), 
September 13, 2002, University of Georgia, Athens, Georgia, USA, and 
to be published in the proceedings.}, 
D.H. Lu$^2$, W. Melnitchouk$^3$, K. Saito$^4$, and A.W. Thomas$^5$
\vspace{1em}}

\address{$^1$Department of Physics and Astronomy, University of Georgia, 
Athens, Georgia 30602, USA \\
$^2$Department of Physics and Zhejiang Institute of Modern Physics, 
Zhejiang University, Hangzhou 310027, China\\
$^3$Jefferson Lab, 12000 Jefferson Avenue, Newport News, VA 23606, USA\\
$^4$Physics Division, Tohoku College of Pharmacy, Sendai 981-8558, Japan\\
$^5$Department of Physics and Mathematical Physics, and 
Special Research Centre for the Subatomic Structure of Matter, 
Adelaide University, SA 5005, Australia
}
\maketitle

%%%%%%%%%%%%%%%%%%%%%%%
\vspace{-7.5cm}
\hfill JLAB-THY-03-03, \hspace{1ex} ADP-03-107/T545
\vspace{7.5cm}
%%%%%%%%%%%%%%%%%%%%%%%

\vspace{1em}
\begin{abstract}
We discuss the electromagnetic form factors, axial 
form factors, and structure functions of a bound nucleon
in the quark-meson coupling (QMC) model. 
Free space nucleon form factors are calculated using
the improved cloudy bag model (ICBM). After describing 
finite nuclei and nuclear matter in the quark-based   
QMC model, we compute the in-medium modification of the 
bound nucleon form factors in the same framework.
Finally, limits on the medium modification of the bound nucleon
$F_2$ structure function are obtained using the calculated in-medium
electromagnetic form factors and local quark-hadron duality.
\end{abstract}

%\vskip 0.5cm
%\pacs{PACS: ........}

%%%%%%%%%%%%%%%%%%%%%%%%%%%%%%%%%%%%%%%
\section{Introduction}
%%%%%%%%%%%%%%%%%%%%%%%%%%%%%%%%%%%%%%%

Partial restoration of chiral symmetry
in a nuclear medium, to which the reduction in the mass of 
a bound nucleon is sometimes ascribed, 
plays a key role in understanding 
the medium modification of bound nucleon (hadron) 
properties.
At relatively high energies and/or temperature and/or densities,
quark and gluon degrees of freedom are expected to be  
efficient in describing physical phenomena according to 
perturbative QCD.
However, it is not at all obvious whether such  
degrees of freedom are indeed necessary or efficient in describing  
low energy nuclear phenomena, such as the static properties of finite nuclei. 
In this article, we demonstrate that the quark degrees of freedom 
do indeed seem to be necessary to understand recent 
polarization transfer measurements in the 
$^4$He($\vec{e},e'\vec{p}$)$^3$H reaction~\cite{HE4,Strauch}, 
which cannot be explained within the best existing treatments of traditional  
physics (solely based on hadronic degrees of freedom).

Over the past few years there has been considerable 
interest in possible changes in 
bound nucleon properties in a nuclear medium.
There is a significant constraint on the possible change in the
radius of a bound nucleon based on $y$-scaling 
--- especially in $^3$He \cite{yscaling}.
On the other hand, the space (time) component of the effective one-body 
axial coupling constant is known to be quenched~\cite{gaspace}
(enhanced~\cite{gatime}) in Gamow-Teller (first-forbidden)   
nuclear $\beta$ decay, and a change in the charge radius
of a bound proton provides a natural suppression of the Coulomb sum rule
\cite{Coulomb}.
One of the most famous nuclear medium effects --- the nuclear EMC effect
\cite{EMC}, or the change in the inclusive deep-inelastic structure
function of a nucleus relative to that of a free nucleon --- has
stimulated theoretical and experimental efforts for almost two decades
now which seek to understand the dynamics responsible for the change in
the quark-gluon structure of the nucleon in the nuclear
environment~\cite{EMCTH}.

%The EMC effect illustrates an inherent difficulty in identifying genuine
%nuclear quark-gluon effects in a background of purely hadronic physics,
%such as conventional nuclear binding and Fermi motion, associated with
%the nuclear bound state.
%Most features of the nuclear to nucleon structure function ratio can be
%(at least qualitatively) understood in terms of conventional nuclear
%physics~\cite{EMC}.
%On the other hand, some of these features can also be attributed to a
%modification of the intrinsic nucleon structure function in medium.

Recently the search for evidence for some modification of nucleon properties in
medium has been extended to electromagnetic form factors, in polarized
$(\vec e, e' \vec p)$ scattering experiments on $^{16}$O~\cite{O16} and
$^4$He~\cite{HE4,Strauch}.
These experiments measured the ratio of transverse to longitudinal
polarization of the ejected protons, which for a free nucleon is
proportional to the ratio of electric to magnetic elastic form factors
\cite{POLTRANS},
\begin{eqnarray}
{ G_E^p \over G_M^p }
&=& - {P'_x \over P'_z} {E_e + E'_e \over 2M_N} \tan(\theta_e/2)\  .
\label{GEpGMp}
\end{eqnarray}
Here $P'_x$ and $P'_z$ are the transverse and longitudinal polarization
transfer observables, $E_e$ and $E'_e$ the incident and recoil electron
energies, $\theta_e$ the electron scattering angle, and $M_N$ the nucleon
mass.
Compared with the traditional cross section measurements,
polarization transfer experiments provide more sensitive tests of
dynamics, especially of any in-medium changes in the form factor ratios.
The feasibility of this technique was first demonstrated in the
commissioning experiment at Jefferson Lab on $^{16}$O~\cite{O16}
at $Q^2=0.8$~GeV$^2$.
In the subsequent experiment at MAMI on $^4$He \cite{HE4} at
$Q^2 \approx 0.4$~GeV$^2$, and at Jefferson Lab at 
$Q^2=0.5,1.0,1.6$ and $2.6$~GeV$^2$, which had much higher statistics, 
the polarization ratio in $^4$He was found to differ by $\approx 10\%$ from
that in $^1$H.

Conventional models using free nucleon form factors and the best
phenomenologically determined optical potentials and bound state wave
functions, as well as relativistic corrections, meson exchange currents,
isobar contributions and final state interactions
\cite{LAGET,KELLY,UDIAS,FOREST}, fail to account for the observed effect
in $^4$He~\cite{HE4,Strauch}.
Indeed, full agreement with the data was only obtained when, in addition
to these standard nuclear corrections, a small change in the structure of
the bound nucleon, which had been estimated within
the quark-meson coupling (QMC)
model~\cite{escatt,He3,QMCB,HeO,Guichon,Guichonf,Saitof}, was taken into
account.
The final analysis~\cite{Strauch} 
seems to favor this scenario even more, 
although the error bars may still be too large to draw a definite conclusion.

%Regardless of the microscopic origin of the nucleon structure
%modification, if there are density dependent effects which modify the
%quark substructure of the nucleon, then these should leave traces in a
%variety of processes and observables, including structure functions and
%form factors.
%
%Of course, one must caution that the study of off-shell nucleon effects is
%hampered with difficulties in unambiguously identifying effects associated
%with nucleon structure deformation \cite{MST}.
%In principle, one can reshuffle strength from off-shell corrections to
%meson exchange currents or interaction terms \cite{OFFSHELL}, so that
%``off-shell effects'' can only be identified after specifying a
%particular form of the interaction of a nucleon with the surrounding
%nuclear medium.
%Nevertheless, within a given model of the nucleus, one can study the
%capacity to {\em simultaneously} describe form factors and structure
%functions as well as static nuclear properties.
%It is in this context that we proceed with the discussion of the possible
%modifications of nucleon properties in the nuclear medium.

On the other hand, there has recently been considerable 
interest in the interplay between
form factors and structure functions in the context of quark-hadron
duality.
As observed originally by Bloom and Gilman \cite{BG}, the $F_2$ structure
function measured in inclusive lepton scattering at low $W$ (where $W$ is
the mass of the hadronic final state) generally follows a global scaling
curve which describes high $W$ data, to which the resonance structure
function averages.
Furthermore, the equivalence of the averaged resonance and scaling
structure functions appears to hold for each resonance region, over
restricted intervals of $W$, so that the resonance--scaling duality
also exists locally.
These findings were dramatically confirmed in recent high-precision
measurements of the proton and deuteron $F_2$ structure function at
Jefferson Lab \cite{JLABF2,JLABPAR}, which demonstrated that local
duality works remarkably well for each of the low-lying resonances,
including surprisingly the elastic, to rather low values of $Q^2$.

In this article we first briefly review how finite nuclei 
and nuclear matter are treated in 
the quark-based QMC model~\cite{Guichonf,Saitof}. We then discuss  
the modification of the electromagnetic and axial form factors
of a bound nucleon in the same model.
Finally, using the concept of quark-hadron 
duality and the calculated bound nucleon electromagnetic form factors, 
we extract the $F_2$ structure function of the bound 
nucleon~\cite{QMCdual}.
To the extent that local duality is a good approximation, the relations
among the nucleon form factors and structure functions 
are model independent, and can in fact be used to test the
self-consistency of the models.
We find that the recent form factor data for a proton bound in 
$^4$He~\cite{HE4,Strauch} place strong constraints on 
the medium modification of
inclusive structure functions at large Bjorken-$x$.
In particular, they appear to disfavor models in which the bulk of the
nuclear EMC effect is attributed to deformation of the intrinsic nucleon
structure off-shell -- see e.g. Ref.~\cite{FS}.

This article is organized as follows.
In Section~II we briefly review the treatment of 
finite nuclei in the QMC model~\cite{Guichonf,Saitof}.
Then, in Section~III, we discuss the in-medium modification of 
bound nucleon electromagnetic form factors in
the QMC model~\cite{escatt,He3,QMCB,HeO}
as inferred from the recent polarization transfer experiments, as well
as that of the axial form factor $G_A(Q^2)$~\cite{GA}. 
%As found in the analysis of the data in Ref.~\cite{HE4,Strauch}, amongst those
%models for which predictions were available, the modifications could only
%be understood within the context of the QMC
%model~\cite{escatt,He3,QMCB,HeO,Guichon,Guichonf,Saitof}.
In Section~IV quark-hadron duality is used to relate the observed form
factor modification to that which would be expected in the deep-inelastic
$F_2$ structure function~\cite{QMCdual}.
Finally, we summarize our findings in Section~V.
%We briefly review the relevant features of Bloom-Gilman duality and
%compare the results of models with and without large medium modifications
%of the intrinsic nucleon structure.

%%%%%%%%%%%%%%%%%%%%%%%%%%%%%%%%%%%%%%%%%%%%%%%%%%%%%%%%%%%%%%%%%%%%%%%%%%%
\section{Finite nuclei and nuclear matter in the QMC model}

In this Section we briefly review the treatment of finite nuclei and
symmetric nuclear matter in the QMC model~\cite{Guichon,Guichonf,Saitof}.
We consider static, spherically symmetric nuclei, and adopt the Hartree,
mean-field approximation,
%
%In this approximation, $\rho NN$ tensor coupling gives
%a spin-orbit force for a nucleon bound
%in a static spherical nucleus, although
%in Hartree-Fock it can give a central force which contributes to
%the bulk symmetry energy~\cite{Guichonf,Saitof}.
%Furthermore, it gives no contribution for nuclear
%matter since the meson fields are independent of position
%and time. 
%
ignoring the $\rho NN$ tensor coupling
as usually done in the Hartree treatment of
quantum hadrodynamics (QHD)~\cite{QHD}
(see Refs.~\cite{Guichonf,Saitof} for discussions  
about the $\rho NN$ tensor coupling).

Using the Born-Oppenheimer approximation, mean-field equations
of motion are derived for a nucleus
in which the quasi-particles moving
in single-particle orbits are three-quark clusters with the quantum numbers
of a nucleon.
A relativistic
Lagrangian density at the hadronic
level can then be constructed~\cite{Guichonf,Saitof},
similar to that obtained in QHD~\cite{QHD},
which produces the same equations of motion
when expanded to the same order in velocity:
\begin{eqnarray}
{\cal L}_{QMC}=& &\overline{\psi}_N(\vec{r})
\left[ i \gamma \cdot \partial
- M_N^{*}(\sigma) - (\, g_\omega \omega(\vec{r})
+ g_\rho \frac{\tau^N_3}{2} b(\vec{r})
+ \frac{e}{2} (1+\tau^N_3) A(\vec{r}) \,) \gamma_0
\right] \psi_N(\vec{r}) \quad \nn \\
-& &\frac{1}{2}[ (\nabla \sigma(\vec{r}))^2 +
m_{\sigma}^2 \sigma(\vec{r})^2 ]
+ \frac{1}{2}[ (\nabla \omega(\vec{r}))^2 + m_{\omega}^2
\omega(\vec{r})^2 ] \nn \\
+& &\frac{1}{2}[ (\nabla b(\vec{r}))^2 + m_{\rho}^2 b(\vec{r})^2 ]
+ \frac{1}{2} (\nabla A(\vec{r}))^2, 
\label{LQMC}
\end{eqnarray}
where $\psi_N(\vec{r})$ 
and $b(\vec{r})$ are respectively the
nucleon and $\rho$
meson (the time component in the third direction of
isospin) fields, while $m_\sigma$, $m_\omega$ and $m_{\rho}$ are
the masses of the $\sigma$, $\omega$ and $\rho$ meson fields.
$g_\omega$ and $g_{\rho}$ are the $\omega$-$N$ and $\rho$-$N$
coupling constants which are related to the corresponding
($u,d$)-quark-$\omega$, $g_\omega^q$, and
($u,d$)-quark-$\rho$, $g_\rho^q$, coupling constants by
$g_\omega = 3 g_\omega^q$ and
$g_\rho = g_\rho^q$~\cite{Guichonf,Saitof}.
(Hereafter, we will denote the light quark flavors by
$q \equiv u,d$.)
The field dependent $\sigma$-$N$ 
coupling strength predicted by the QMC model, $g_\sigma(\sigma)$, 
which is related to the Lagrangian density of 
Eq.~(\ref{LQMC}) at the hadronic level, is defined by:
\bg
M_N^{*}(\sigma) &\equiv& M_N - g_\sigma(\sigma)
\sigma(\vec{r})\ .  
\label{MN}
\en
%
% where $M_N$ is the free nucleon mass. % ... M_N defined earlier ...
Note that the dependence of these coupling strengths on the applied
scalar field must be calculated self-consistently at the quark level
\cite{Guichonf,Saitof}.
%
% More explicit expression for 
% $g_\sigma(\sigma)$ will be given later.
%
{}From the Lagrangian density in
Eq.~(\ref{LQMC}), a set of
equations of motion for the nuclear system can be obtained:
\begin{eqnarray}
& &[i\gamma \cdot \partial -M^*_N(\sigma)-
(\, g_\omega \omega(\vec{r}) + g_\rho \frac{\tau^N_3}{2} b(\vec{r})
 + \frac{e}{2} (1+\tau^N_3) A(\vec{r}) \,)
\gamma_0 ] \psi_N(\vec{r}) = 0, 
\label{eqdiracn}\\
& &(-\nabla^2_r+m^2_\sigma)\sigma(\vec{r}) =
- [\frac{\partial M_N^*(\sigma)}{\partial \sigma}]\rho_s(\vec{r})
\equiv g_\sigma C_N(\sigma) \rho_s(\vec{r}), 
\label{eqsigma}\\
& &(-\nabla^2_r+m^2_\omega) \omega(\vec{r}) =
g_\omega \rho_B(\vec{r}), \label{eqomega}\\ 
& &(-\nabla^2_r+m^2_\rho) b(\vec{r}) =
\frac{g_\rho}{2}\rho_3(\vec{r}),
\label{eqrho}\\
& &(-\nabla^2_r) A(\vec{r}) =
e \rho_p(\vec{r}), \label{eqcoulomb}
\end{eqnarray}
where, $\rho_s(\vec{r})$, $\rho_B(\vec{r})$, $\rho_3(\vec{r})$ and
$\rho_p(\vec{r})$ are the scalar, baryon, third component of isovector,
and proton densities at position $\vec{r}$ in 
the nucleus~\cite{Guichonf,Saitof}.
On the right hand side of Eq.~(\ref{eqsigma}),
$- [{\partial M_N^*(\sigma)}/{\partial \sigma}] =
g_\sigma C_N(\sigma)$, where $g_\sigma \equiv g_\sigma (\sigma=0)$,  
is a new, and characteristic feature of QMC
beyond QHD~\cite{QHD}.
The effective mass for the nucleon, $M_N^*$, is defined by:
\begin{equation}
\frac{\partial M_N^*(\sigma)}{\partial \sigma}
= - n_q g_{\sigma}^q \int_{bag} d^3 x
\ {\overline \psi}_q(\vec{x}) \psi_q(\vec{x})
\equiv - n_q g_{\sigma}^q S_N(\sigma) = - \frac{\partial}{\partial \sigma}
\left[ g_\sigma(\sigma) \sigma \right],
\label{gsigma}
\end{equation}
where $n_q$ is the number of light quarks ($u$ and $d$), and  
the MIT bag model quantities and
the in-medium bag radius satisfying the
mass stability condition are given by~\cite{Guichon,Guichonf,Saitof}:
\begin{eqnarray}
& &M_N^*(\sigma) =
\sum_{q=u,d}\frac{n_q\Omega^*_q -  z_N}{R_N^*}
+ \frac{4}{3}\pi ({R_N^*})^3 B\ ,
\label{MC}\\
& &S_N(\sigma) = 
\left[\Omega_q^*/2+m_q^*R_N^*(\Omega_q^*-1)\right]/
\left[\Omega_q^*(\Omega_q^*-1)+ m_q^*R_N^*/2\right]\ ,
\label{Ssigma}\\
& &
\Omega_q^* = \sqrt{x_q^2 + (R_N^* m_q^*)^2},\hspace{2ex} 
m_q^* = m_q - g_{\sigma}^q \sigma (\vec{r})\ , 
\label{Omega}\\
& & \left. dM^*_{N}/dR_{N}
\right|_{R_{N} = R^*_{N}} = 0\ ,
\label{bagradii}
\end{eqnarray}
where $g_\sigma^q$ is the quark-$\sigma$ meson coupling constant.
Here, the MIT bag model quantities are calculated in a local
density approximation using the
spin and spatial part of the wave functions,
$\psi_q (x) = N_q e^{- i \epsilon_q t / R_N^*}\psi_q (\vec{x})$,
where $N_q$ is the normalization factor.
The wave functions $\psi_q (x)$ satisfy the Dirac equations for
the quarks in the nucleon bag
centered at position $\vec{r}$ in the nucleus 
($|\vec{x} - \vec{r}|\le R_N^*$~\cite{Guichonf,Saitof}):
\begin{eqnarray}
\left[ i \gamma \cdot \partial_x -
(m_q - V^q_\sigma(\vec{r}))
\mp \gamma^0
\left( V^q_\omega(\vec{r}) \pm
\frac{1}{2} V^q_\rho(\vec{r})
\right) \right]
\left( \begin{array}{c} \psi_u(x) \\
\psi_d(x) \\ \end{array} \right) &=& 0\ ,
\label{Diracud}
\end{eqnarray}
where we approximate the constant, mean meson fields
within the bag and neglect the Coulomb force.
The constant, mean-field potentials within the bag centered at
$\vec{r}$
are defined by $V^q_\sigma(\vec{r}) \equiv g^q_\sigma \sigma(\vec{r})$,
$V^q_\omega(\vec{r}) \equiv g^q_\omega \omega(\vec{r})$ and
$V^q_\rho(\vec{r}) \equiv g^q_\rho b(\vec{r})$.
%
% with $g^q_\sigma$, $g^q_\omega$ and
% $g^q_\rho$ the corresponding quark-meson coupling constants.
%
The eigenenergies
in units of $1/R_N^*$ are given by:
%%%%%%%%%
\bge
\left( \begin{array}{c}
\epsilon_u \\
\epsilon_d
\end{array} \right)
= \Omega_q^* \pm R_N^* \left(
V^q_\omega(\vec{r})
\pm \frac{1}{2} V^q_\rho(\vec{r}) \right).\,\,
\label{energy}
\ene
%%%%%%%%%
In Eqs.~(\ref{MC})~-~(\ref{bagradii}), 
$z_N$, $B$, $x_q$, and $m_q$ are the parameters
for the sum of the c.m.
and gluon fluctuation effects,
bag pressure, lowest eigenvalues for the quark $q$, 
and the corresponding current quark masses, respectively.
$z_N$ and $B$ are fixed by fitting the nucleon mass
in free space. We use the current quark masses $m_{q=u,d} = 5$ MeV,
and obtained $z_N = 3.295$ and $B = (170.0$ MeV$)^4$ by choosing the
bag radius for the nucleon in free space $R_N = 0.8$~fm.
The parameters at the hadronic level, which are already fixed by the study of
nuclear matter and finite nuclei~\cite{Saitof},
are as follows: $m_\omega = 783$ MeV, $m_\rho = 770$ MeV, $m_\sigma = 418$ MeV,
$e^2/4\pi = 1/137.036$, $g^2_\sigma/4\pi = 3.12$, $g^2_\omega/4\pi = 5.31$
and $g^2_\rho/4\pi = 6.93$.
The sign of $m_q^*$ in the nucleus in Eq.~(\ref{Omega})
reflects nothing but the strength
of the attractive, negative scalar potential,
and thus the naive interpretation of the mass for a physical particle,
which is positive, should not be applied.

At the hadronic level, the entire information
on the quark dynamics is condensed into the effective coupling
$C_N(\sigma)$ of Eq.~(\ref{eqsigma}).
Furthermore, when $C_N(\sigma) = 1$, which corresponds to
a structureless nucleon, the equations of motion
given by Eqs.~(\ref{eqdiracn})-(\ref{eqcoulomb})
can be identified with those derived
from QHD~\cite{QHD},
except for the terms arising from the tensor coupling and the non-linear
scalar field interaction introduced beyond naive QHD.
%%%%%%%%%%%%%%%%%%%%%%%%%%%%%%%%%%%%%%%
\begin{figure}[htb]
\begin{center}
\epsfig{file=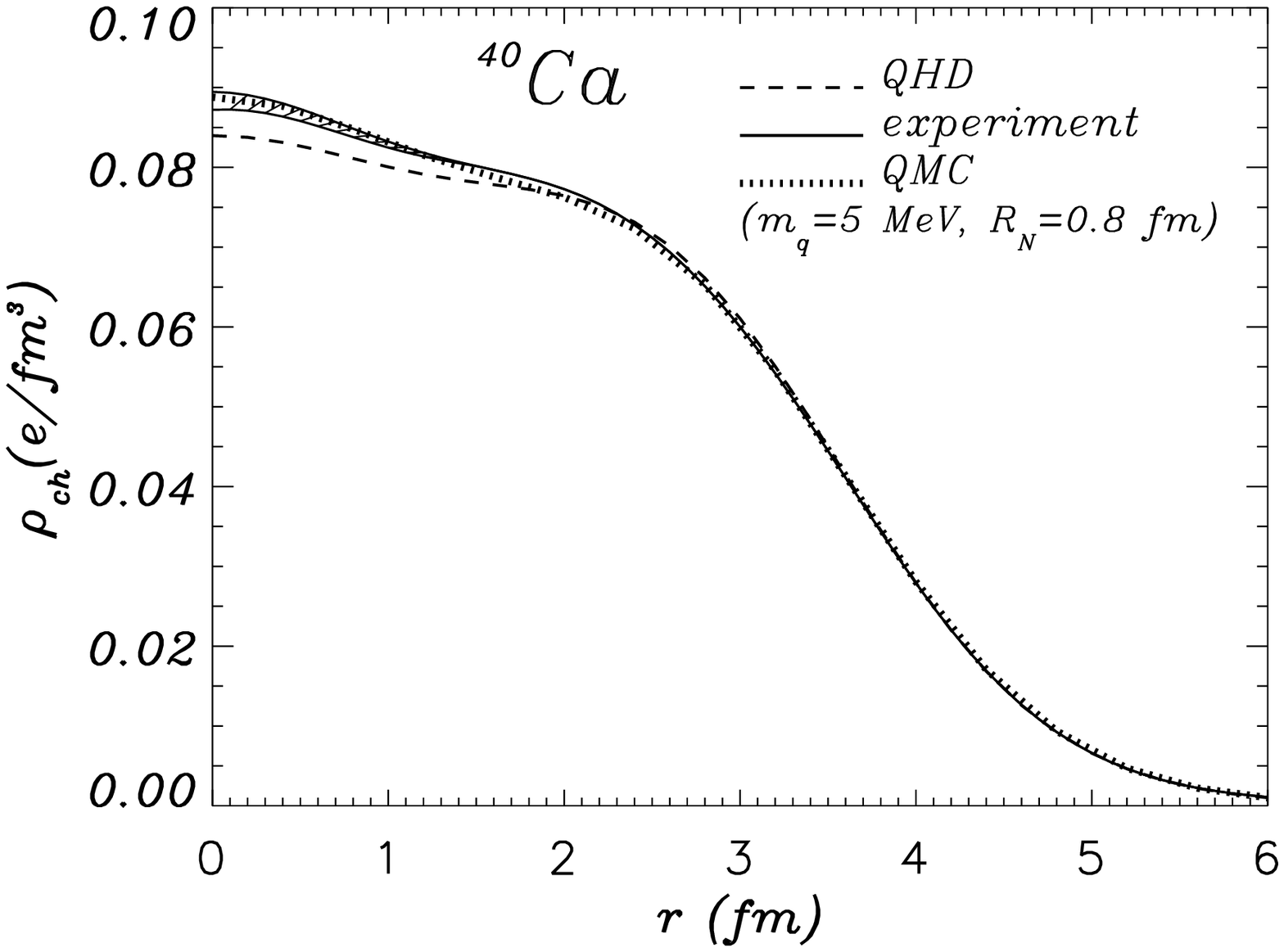,height=6.5cm,width=8.0cm}
\epsfig{file=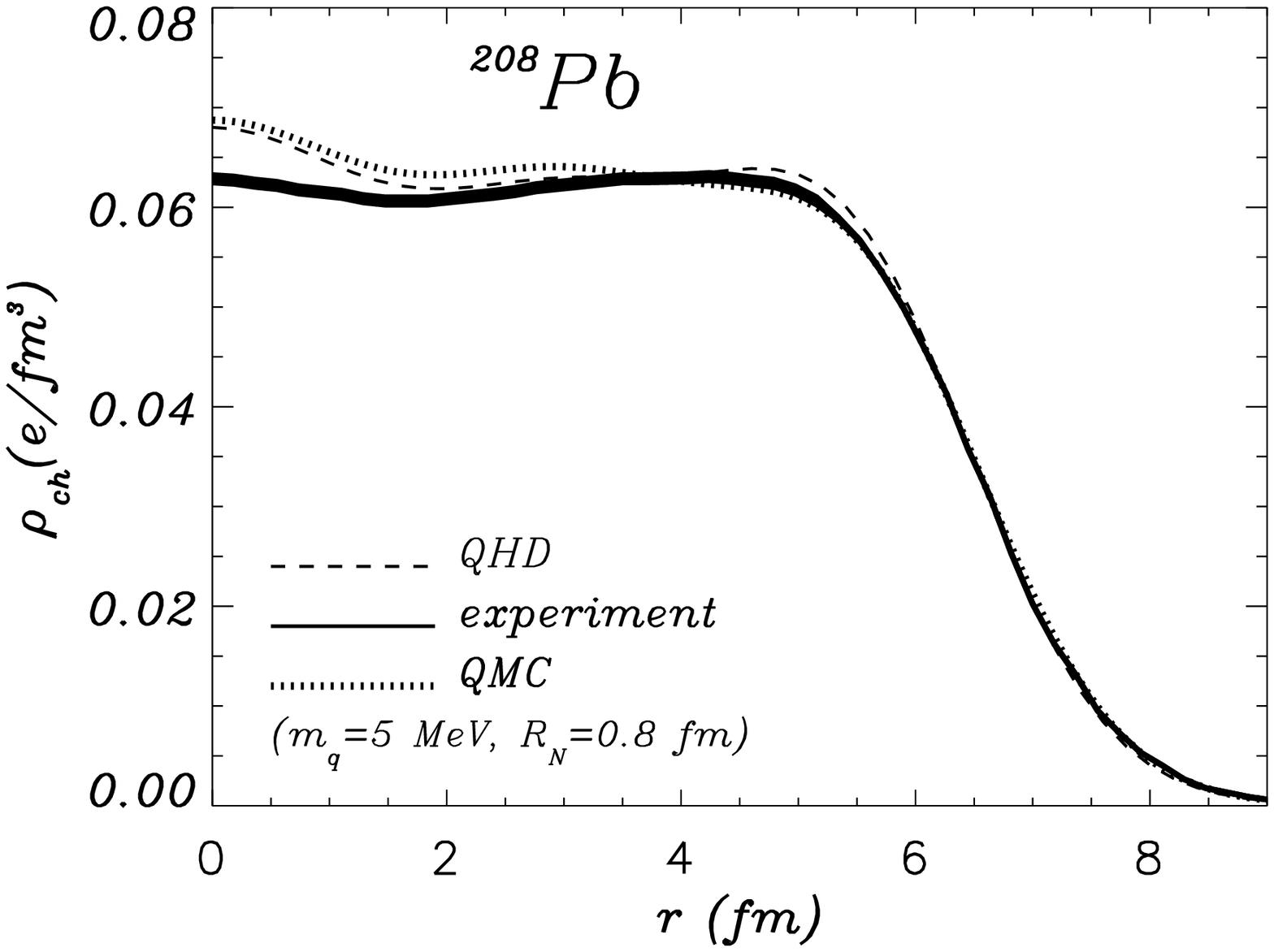,height=6.5cm,width=8.0cm}
%\vspace{1em}
\caption{Charge density distributions for $^{40}$Ca and $^{208}$Pb 
calculated in the QHD~\protect\cite{QHD} and QMC~\protect\cite{Saitof}
models.
\label{charge}
}
\end{center}
\end{figure}
%%%%%%%%%%%%%%%%%%%%%%%%%%%%%%%%%%%%%%%

\begin{figure}[htb]
\begin{center}
\epsfig{file=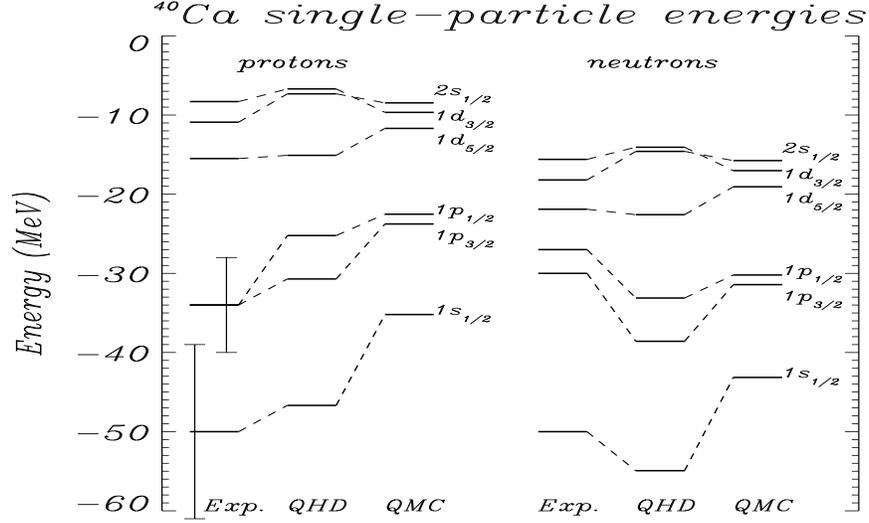,height=8.0cm,width=13cm}
\vspace{-0.5cm}
\caption{Energy spectrum for $^{40}$Ca~\protect\cite{Saitof}
in the QMC model
compared with experiment ($Exp.$), and that of QHD~\protect\cite{QHD}.
\label{speca}
}
\vspace{-0.5cm}
\end{center}
\end{figure}

\begin{figure}[htb]
\begin{center}
\epsfig{file=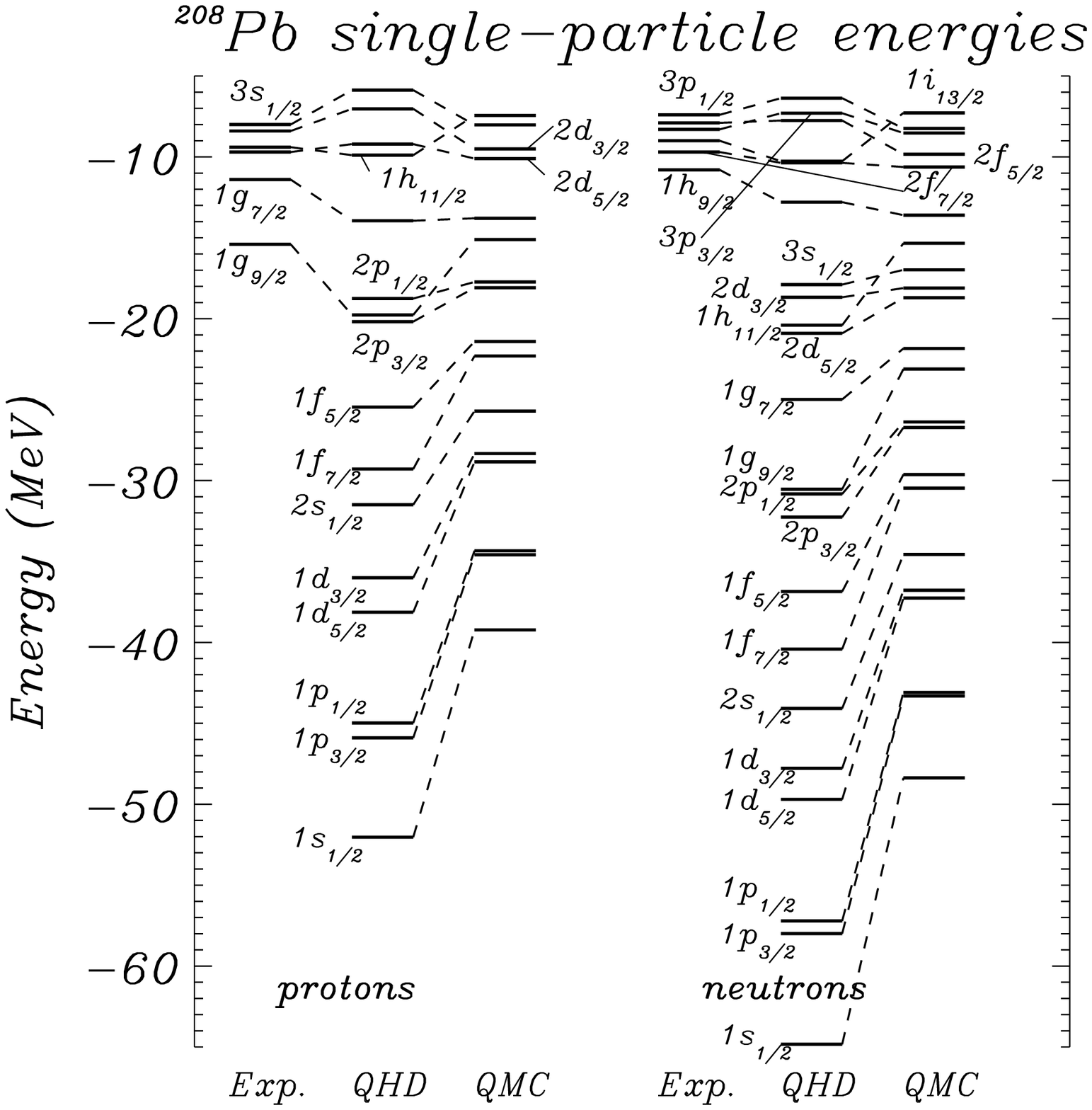,height=11cm,width=15cm}
\vspace{-0.5cm}
\caption{Same as Fig.~\protect\ref{speca} (for $^{209}$Pb).
\label{spepb}
}
\end{center}
\end{figure}

As examples, we show in Fig.~\ref{charge}
the charge density distributions calculated
for $^{40}$Ca and $^{208}$Pb,
and also the energy spectra obtained for $^{40}$Ca and $^{208}$Pb
in Figs.~\ref{speca} and~\ref{spepb}~\cite{Saitof}, respectively.

Next, we consider the limit of infinite symmetric nuclear 
matter~\cite{Guichon,Guichonf,Saitof}.
In this limit all meson fields
become constant, and we denote the mean-values of the 
$\omega$ and $\sigma$ fields by
$\overline{\omega}$ and $\overline{\sigma}$.
Then, equations for the $\overline{\omega}$ and 
self-consistency condition for the $\overline{\sigma}$  
%and effective nucleon mass, $M_N^*$,
are given by~\cite{Guichon,Guichonf,Saitof},
\bg
\overline{\omega} &=& \frac{4}{(2\pi)^3}\int d^3 k\, \theta (k_F - k) 
= \frac{g_\omega}{m_\omega^2} \frac{2k_F^3}{3\pi^2}
= \frac{g_\omega}{m_\omega^2} \rho_B,  
\label{omega_bar}\\
\overline{\sigma}&=&\frac{g_\sigma }{m_\sigma^2}C_N(\overline{\sigma})
\frac{4}{(2\pi)^3}\int d^3 k\, \theta (k_F - k)
\frac{M_N^{*}(\overline{\sigma})}
{\sqrt{M_N^{* 2}(\overline{\sigma})+\vec{k}^2}} 
= \frac{g_\sigma }{m_\sigma^2}C_N(\overline{\sigma}) \rho_s, 
\label{scc}
\en
where $g_{\sigma} = 3 g^q_\sigma S_N(0)$ (see Eq.~(\ref{Ssigma})), $k_F$ is the
Fermi momentum, $\rho_B$ and $\rho_s$ are the baryon and scalar 
densities, respectively. 
%and $C_N(\overline{\sigma})$ is now the constant value of $C_N$ in the
%scalar field.
Note that $M_N^* (\overline{\sigma})$ 
in Eq.~(\ref{scc}) must be calculated
self-consistently in the MIT bag model through
Eqs.~(\ref{gsigma})--(\ref{Diracud}) for a given baryon density.
This self-consistency equation for the $\overline{\sigma}$
is the same as that in QHD, except that in the latter model one has
$C_N(\overline{\sigma})=1$~\cite{QHD}.
%%%%%%%%%%%%%%%%%%%%%%%%%%%%%%%%%%%%%%%
%%%%%%%%%%%%%%%%%%%%%%%%%%%%%%%%%%%%%%%
%\vspace{-0.5cm}
%%%%%%%%%%%%%%%%%%%%%%%%%%%%%%%%%%%%%%%
%%%%%%%%%%%%%%%%%%%%%%%%%%%%%%%%%%%%%%%
%%%%%%%%%%%%%%%%%%%%%%%%%%%%%%%%%%%%%%%%%%%%%%%%%%%%%%%%%%%%%%%%%%%%%%%%%%%
\section{Nuclear Medium Modification of Form Factors}

In this Section we outline the medium modification of the electromagnetic
form factors of the nucleon, as suggested in the recent 
polarization transfer measurements in the
$^4$He($\vec{e},e'\vec{p}$)$^3$H reaction~\cite{HE4,Strauch}.
The first data were analyzed in Ref.~\cite{HE4} 
using a variety of models, nonrelativistic and
relativistic, based on conventional nucleon-nucleon potentials and
well-established bound state wave functions, including corrections from
meson exchange currents, final state interaction and other effects
\cite{LAGET,KELLY,UDIAS,FOREST}.
%The conventional models with the free nucleon form factors could produce
%a deviation of at most one half of a percent in the nuclear transverse to
%longitudinal ratio, $P'_x/P'_z$, compared with that in hydrogen, although
%spinor distortions in fully relativistic calculations were found to
%produce an effect of order 2--5\% \cite{UDIAS}.
The observed deviation, which was of order 10\%, could only be explained
by supplementing the conventional nuclear description with the effects
associated with the medium modification of the nucleon internal structure 
calculated by the QMC model~\cite{escatt,He3,QMCB,HeO}.
%Even though the effect is currently only at the level of 1--2
%standard deviations, it is of considerable interest and importance as
%the first relatively model independent indication of a change in the
%internal structure of the nucleon in a nuclear environment.

%As a result, the internal structure of the bound nucleon is modified by
%the surrounding nuclear medium.

%Because the average nuclear densities for all existing stable nuclei
%heavier than deuterium lie in the range
%${1\over 2} \rho_0 \alt \rho \alt \rho_0$, where
%$\rho_0 = 0.15$~fm$^{-3}$ is the normal nuclear matter density,
%we consider two specific nuclear densities ($\rho = {1\over 2} \rho_0$
%and $\rho = \rho_0$) to give the upper and lower bounds for the change
%of the electromagnetic form factors (and structure functions at large $x$)
%of the bound nucleon.
%Furthermore, for the isoscalar $^4$He and $^{16}$O nuclei we neglect the
%tiny amount of charge symmetry breaking (due to the Coulomb force and the
%$u$ and $d$ current quark mass differences).

\begin{figure}[htb]
\begin{center}
\epsfig{file=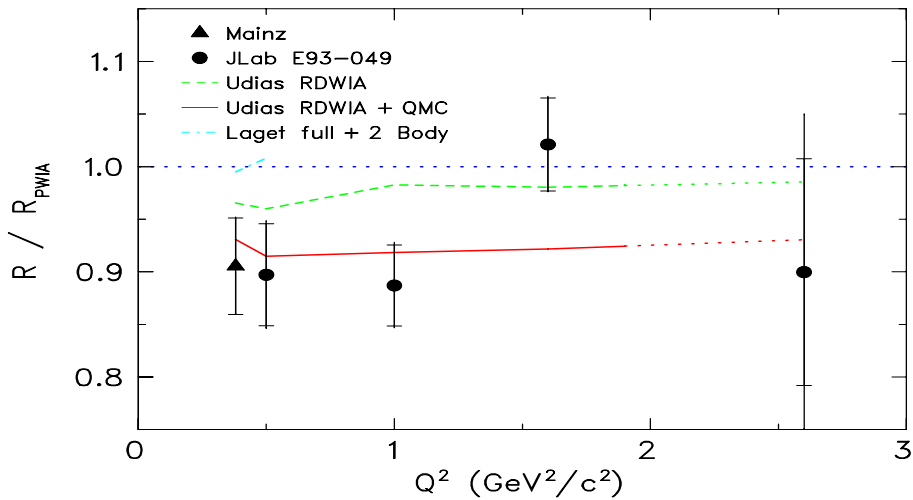,angle=90,height=6.5cm,width=12cm}
\vspace{1em}
\caption{Super ratio $R/R_{PWIA}$, as a function of $Q^2$, taken from
Ref.~\protect\cite{Strauch}. See caption of Fig.~1 in 
Ref.~\protect\cite{Strauch} for detailed explanations. 
\label{JlabQMC}}
\end{center}
\end{figure}

%
In Fig.~\ref{JlabQMC} we show the ``super ratio", $R/R_{PWIA}$, which was 
made for the final analysis of the polarization transfer measurements 
on $^4$He~\cite{Strauch}.
Here, $R_{PWIA}$ stands for the prediction
based on the relativistic plane-wave impulse approximation (PWIA),
and the measured ratio $R$ is defined by:
\bge
R = \frac{(P_x'/P'_z)_{^4He}}{(P_x'/P'_z)_{^1H}}\,\, .
\label{sratio}
\ene
 
%%%%%%%%%%%%%%%%%%%%%%%%%%%%%%%%%%%%%%%
%%%%%%%%%%%%%%%%%%%%%%%%%%%%%%%%%%%%%%%

In Fig.~\ref{JlabQMC}, the modification of electromagnetic form factors 
of the bound nucleon calculated in the QMC model~\cite{escatt,He3,QMCB,HeO}
(the solid line denoted by ``Udias RDWIA + QMC") uses the improved cloudy
bag model (ICBM)~\cite{CBM,DING} for the free nucleon form factors.
The ICBM~\cite{DING} includes a Peierls-Thouless projection to account for
center of mass and recoil corrections, and a Lorentz contraction of the
internal quark wave functions.

\begin{figure}[htb]
\begin{center}
\epsfig{file=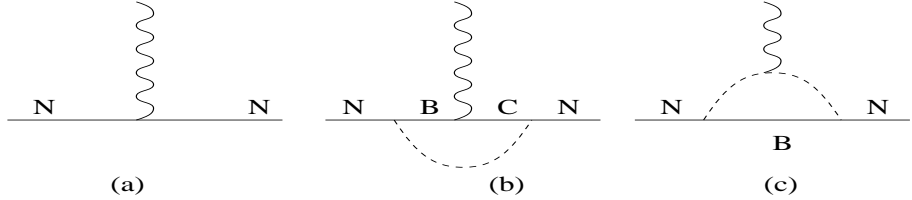,height=4.5cm,width=12cm}
\vspace{1em}
\caption{Diagrams included in the calculation of free space electromagnetic 
form factors in the ICBM~\protect\cite{DING}.
The intermediate baryon states $B$ and $C$ are restricted to the 
$N$ and $\Delta$.
\label{CBMdiagram}}
\end{center}
\end{figure}

The electromagnetic current is given by the sum of the contributions
from the quark core and the pion cloud,
\begin{eqnarray}
\label{current}
j^\mu(x) = \sum_q Q_q e \overline{\psi}_q(x) \gamma^\mu \psi_q(x)
         - i e [ \pi^\dagger(x) \partial^\mu \pi(x)
                -\pi(x) \partial^\mu \pi^\dagger(x) ]\ ,
\end{eqnarray}
where $Q_q$ is the charge operator for a quark flavor $q$, and $\pi(x)$
destroys a negatively charged (or creates a positively charged) pion.
Relevant diagrams included in the calculation of free space   
electromagnetic form factors 
are depicted in Fig.~\ref{CBMdiagram}~\cite{DING}.

%%%%%%%%%%%%%%%%%%%%%%%%%%%%%%%%%%%%%%%
\vspace{-1.5cm}
%%%%%%%%%%%%%%%%%%%%%%%%%%%%%%%%%%%%%%%

In the Breit frame the quark core contribution to the electromagnetic
form factors of the bound nucleon is given by~\cite{escatt,He3,QMCB,HeO}:
\begin{mathletters}
\label{GEMQMC}
\begin{eqnarray}
G_E^*(Q^2)
&=& \eta^2\ G^{\rm sph\, *}_E(\eta^2 Q^2)\ ,\label{GEmedium}        \\
G_M^*(Q^2)
&=& \eta^2\ G^{\rm sph\, *}_M(\eta^2 Q^2)\ ,\label{GMmedium}
\end{eqnarray}
\end{mathletters}%
where $Q^2 \equiv -q^2 = \vec{q}^{\,2}$, and the scaling factor
$\eta = (M_N^*/E_N^*)$, with $E_N^*=\sqrt{{M_N^*}^2 + Q^2/4}$ the energy
and $M_N^*$ the mass of the nucleon in medium.
$G_{E,M}^{\rm sph\, *}$ are the form factors calculated with the
static spherical bag wave function,
\begin{mathletters}
\begin{eqnarray}
G_E^{\rm sph\, *}(Q^2)
&=& { 1 \over D }
    \int\! d^3r\ j_0(Qr)\ f_q(r)\ K(r)\ ,               
\label{GE}\\
G_M^{\rm sph\, *}(Q^2)
&=& { 1 \over D }
    { 2 M^*_N \over Q }
    \int\! d^3r\ j_1(Qr)\ \beta_q^*\
    j_0(x_q r/R_N^*)\ j_1(x_q r/R_N^*)\ K(r)\ .
\label{GM}
\end{eqnarray}
\end{mathletters}%
%
% and are subject to $Q \to \eta Q$:
%
Here
$f_q(r) = j_0^2(x_q r/R_N^*) + \beta_q^{*2}\ j_1^2(x_q r/R_N^*)$,
where $R_N^*$ is the nucleon bag radius in medium, 
$x_q$ the lowest eigenfrequency, and 
$\beta_q^{*2} = (\Omega_q^* - m_q^* R_N^*)/(\Omega_q^* + m_q^* R_N^*)$, 
with
$\Omega_q^* = \sqrt{ x_q^2 + (m_q^* R_N^*)^2 }$ and 
$m_q^* = m_q - g^q_\sigma \sigma$ (see also Section~II).
The recoil function
$K(r) = \int\! d^3x \, f_q(\vec x) f_q(-\vec x - \vec r)$
accounts for the correlation of the two spectator quarks, and
$D = \int\! d^3r\ f_q(r)\ K(r)$ is the normalization factor.
The scaling factor $\eta$ in the argument of $G_{E,M}^{\rm sph\, *}$ arises
from the coordinate transformation of the struck quark, and the prefactor
in Eqs.~(\ref{GEMQMC}) comes from the reduction of the integral measure of
the two spectator quarks in the Breit frame.

The contribution from the pionic cloud is calculated along the lines of
Ref.~\cite{escatt,He3,QMCB,HeO}.
Although the pion mass would be slightly smaller in the medium than in
free space, we use $m_\pi^*= m_\pi$, which is consistent with chiral 
expectations and phenomenological constraints.
Furthermore, since the $\Delta$ isobar is treated on the same footing as
the nucleon in the CBM, and because it contains three ground state light
quarks, its mass should vary in a similar manner to that of the nucleon
in the QMC model.
As a first approximation we therefore take the in-medium and free space
$N$--$\Delta$ mass splittings to be approximately equal,
$M_\Delta^* - M_N^* \simeq M_\Delta - M_N$.

%Including both the quark core and pion cloud contributions, the electric
%and magnetic form factors of the free and bound nucleons were calculated
%in Ref.\cite{escatt,He3,QMCB,HeO}.
%One finds that the modification of the bound nucleon form factors is
%1--2\% for the magnetic and of order 8\% for the electric form factor,
%respectively, at normal nuclear matter density ($\rho = \rho_0$), and for
%$Q^2 \simeq 0.3$~GeV$^2$, when all form factors are normalized to unity at
%$Q^2=0$.
%
%Of course, in the present analysis the absolute value of the proton
%magnetic form factor at $Q^2=0$ (the magnetic moment), which is enhanced
%in medium, also plays an important role -- as it did in the analysis of
%polarized $(\vec e, e'\vec p)$ scattering experiments.
%
%The values of the current quark masses, $m_q \equiv m_u = m_d = 5$~MeV,
%and the nucleon bag radius in free space, $R = 0.8$~fm, are the same as
%those used in the earlier calculations which reproduce nuclear saturation
%properties, and which produced the good agreement with the form factor
%data in Ref.\cite{HE4}.
%None of the results for nuclear properties, however, depend strongly on
%the choice of parameters once the quark-meson coupling constants are
%fixed to reproduce the nuclear saturation properties.
%As shown in Ref.\cite{Saitof}, the dependence of the properties of finite
%nuclei on $m_q$ and $R$ is relatively weak.

\begin{figure}[htb]
\begin{center}
\hspace*{-1cm}
\epsfig{file=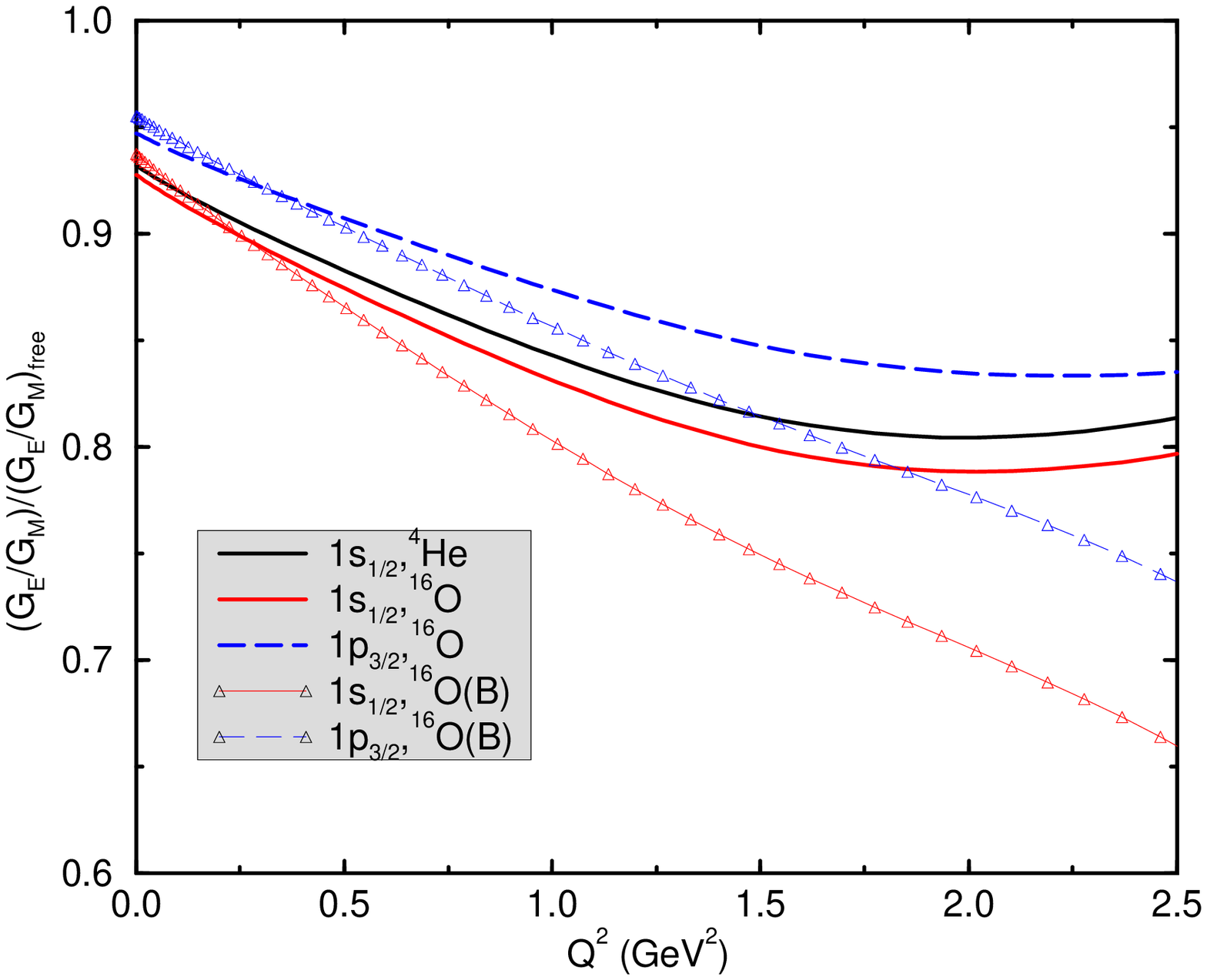,height=7.0cm,width=9.0cm}\hspace{-0.5cm}
\epsfig{file=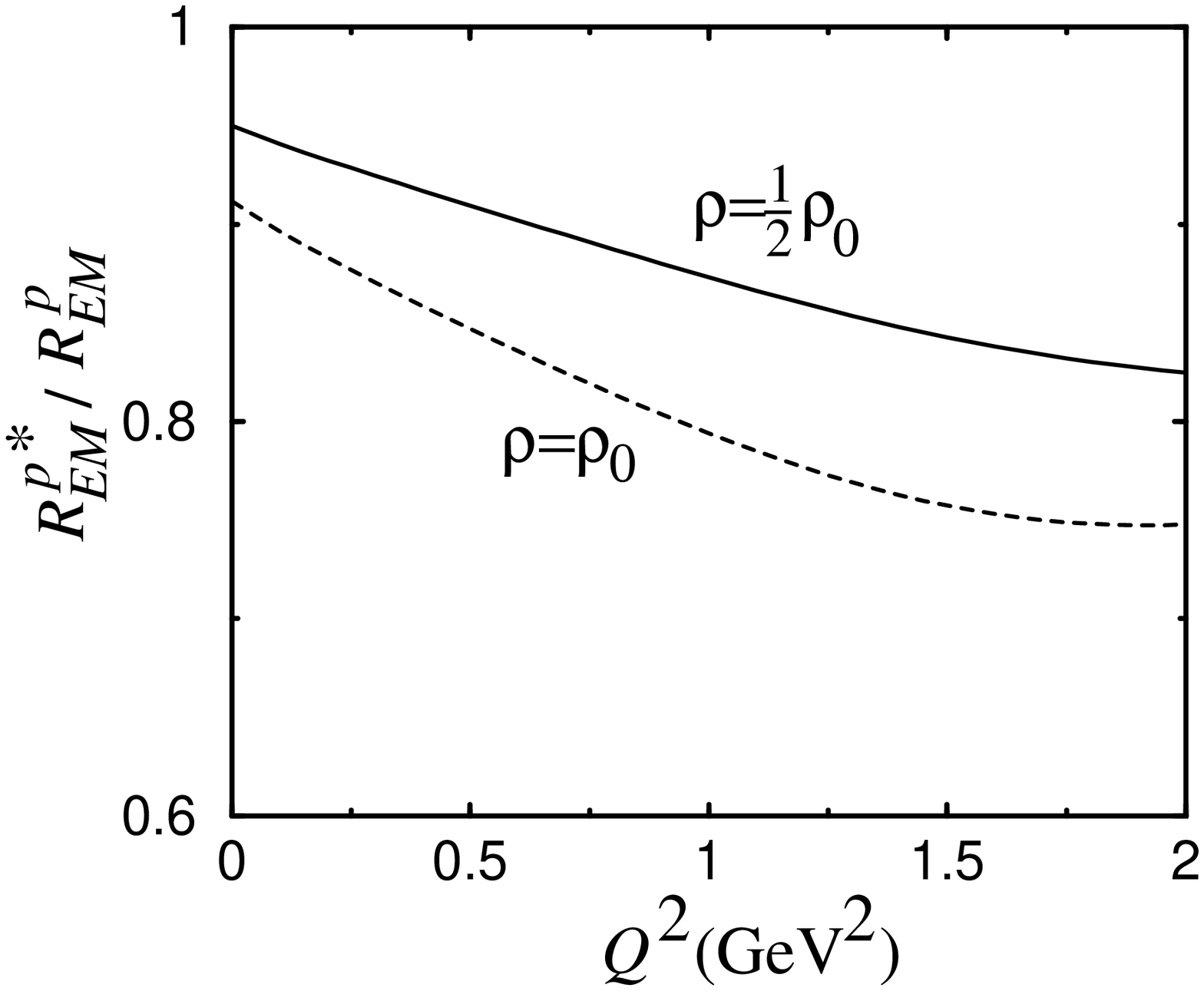,height=6.5cm,width=8.0cm}
\vspace{1em}
\caption{The change in the ratio of the electric to magnetic form factors 
from free to bound protons,   
$R^{p*}_{EM}/R^p_{EM} = (G^{p*}_E/G^{p*}_M)/(G^{p}_E/G^{p}_M)$,
for $^4$He ($1s_{1/2}$ state) and $^{16}$O ($1s_{1/2}$ and $1s_{3/2}$
states) (left panel)~\protect\cite{HeO} and nuclear
matter (right panel)~\protect\cite{QMCdual}. $^{16}$O(B) stands for the results
which allow changes of the bag constant according to the nuclear density
($\rho_0 = 0.15$~fm$^{-3}$~\protect\cite{QMCB}).
\label{REM}
}
\end{center}
\end{figure}

The change in the ratio of the electric to magnetic form factors of the
proton from free to bound, 
\begin{eqnarray}
R^{p\, *}_{EM}(Q^2)\left/ R^{p}_{EM}(Q^2)\right. 
&=& \left({ G_E^{p\, *}(Q^2) \over G_M^{p\, *}(Q^2) }\right)\left/
\left({ G_E^p(Q^2) \over G_M^p(Q^2) }\right)\right. , 
\end{eqnarray}
is illustrated in Fig.~\ref{REM} for 
$^4$He, $^{16}$O (left panel) and for nuclear matter densities, 
$\rho = \rho_0$ and $\rho = {1 \over 2} \rho_0$ (right panel) with 
$\rho_0 = 0.15$~fm$^{-3}$.
%%%%%%%%%%%%%%%%%%%%%%%%%%%%%%%%%%%%%%%
%%%%%%%%%%%%%%%%%%%%%%%%%%%%%%%%%%%%%%%
Because the average nuclear densities for all existing stable nuclei
heavier than deuterium lie in the range
${1\over 2} \rho_0 \alt \rho \alt \rho_0$, 
we consider these two specific nuclear densities
to give the upper and lower bounds for the change
of the electromagnetic form factors (and structure functions at large $x$)
of the bound nucleon.
We emphasize that in the present analysis the absolute value of the proton
magnetic form factor at $Q^2=0$ (the magnetic moment), which is enhanced
in medium, plays an important role -- as it did in the analysis of
polarized $(\vec e, e'\vec p)$ scattering experiments.

Because of charge conservation, the value of $G_E^p$ at $Q^2=0$ remains
unity for any $\rho$.
On the other hand, the proton magnetic moment is enhanced in the nuclear
medium, increasing with $\rho$, so that $R^{p\, *}_{EM} < R^p_{EM}$ at
$Q^2=0$.
In fact, the electric to magnetic ratio is $\sim 5\%$ smaller in medium
than in free space for $\rho={1\over 2}\rho_0$, and $\sim 10\%$
smaller for $\rho=\rho_0$.
The effect increases with $Q^2$ out to $\sim 2$~GeV$^2$, where the
(bound/free) ratio deviates by $\sim 20\%$ from unity.

%On the other hand, because nuclear density is not uniform throughout the
%nucleus, the $\approx$ 20\% change in the form factors produces only
%a few \% effect in the polarization ratio \cite{HE4}.
%The experiment not only probes the central region where $\rho$ is
%maximal, but also outer regions where $\rho$ is much smaller, so that
%integration over the entire nucleus dilutes the effect.
%Nevertheless, a form factor modification of this order of magnitude is
%needed to explain the observed effect \cite{HE4}.
%In the next section we examine the implications of the modification of
%the form factors for the medium modification of structure functions at
%large $x$.

The extension to 
the in-medium modification of the bound nucleon 
axial form factor $G_A^*(Q^2)$ 
can be made in a straightforward manner~\cite{GA}.
Since the induced pseudoscalar form factor, $G_P(Q^2)$,
is dominated by the pion pole, and can be derived using
the PCAC relation~\cite{CBM}, we do not discuss it here.
The relevant axial current operator is then simply given by 
\begin{eqnarray}
A^\mu_a(x) &=& \sum_q \overline{\psi}_q(x) \gamma^\mu\gamma_5
{\tau_a\over 2} \psi_q(x) \theta(R-r), 
\label{Acurrent}
\end{eqnarray}
where $\psi_q(x)$ is the quark field operator for flavor $q$.

\begin{figure}[htb]
\begin{center}
%\hspace*{-1cm}
\epsfig{file=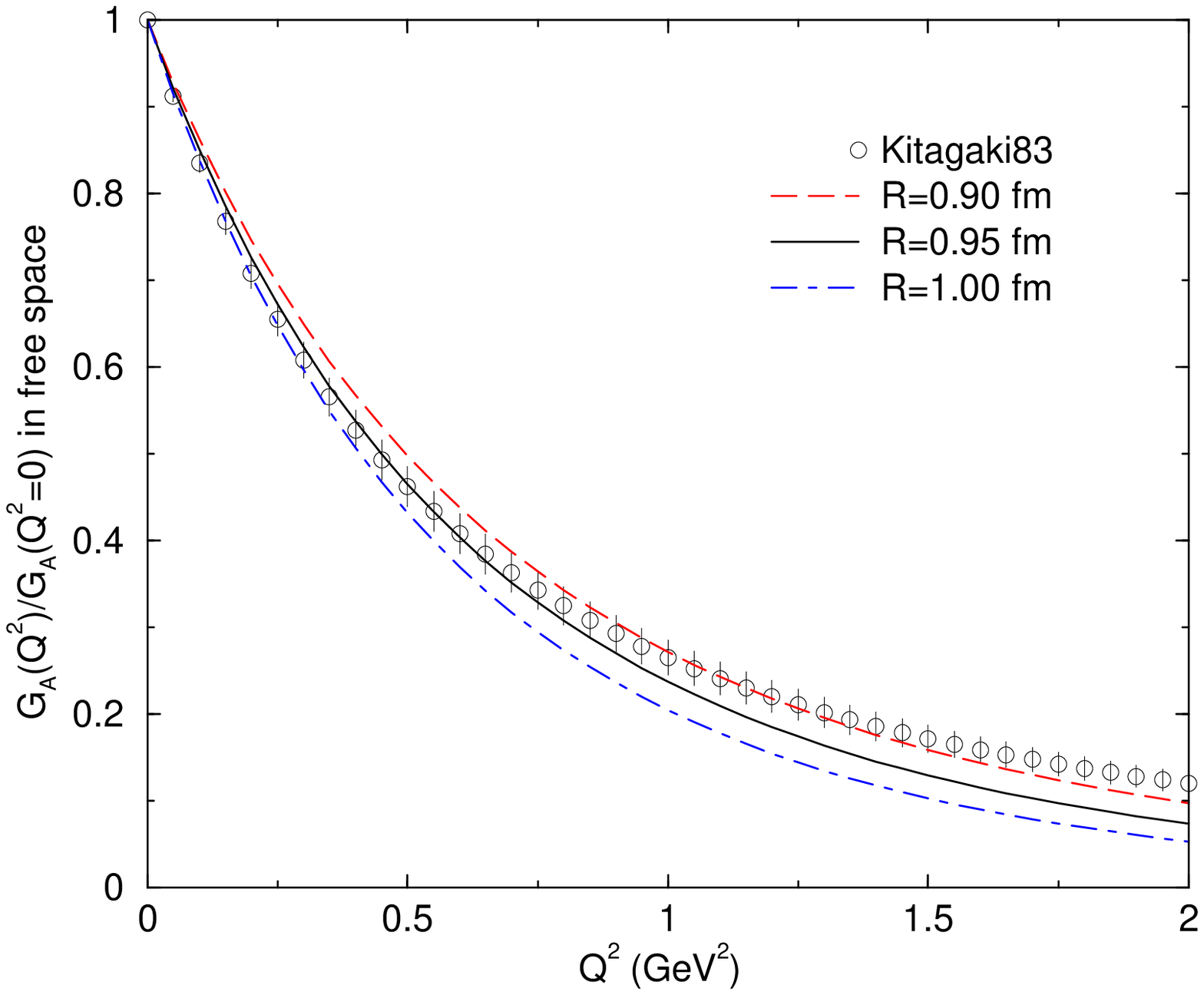,height=6.5cm,width=8.0cm}
\epsfig{file=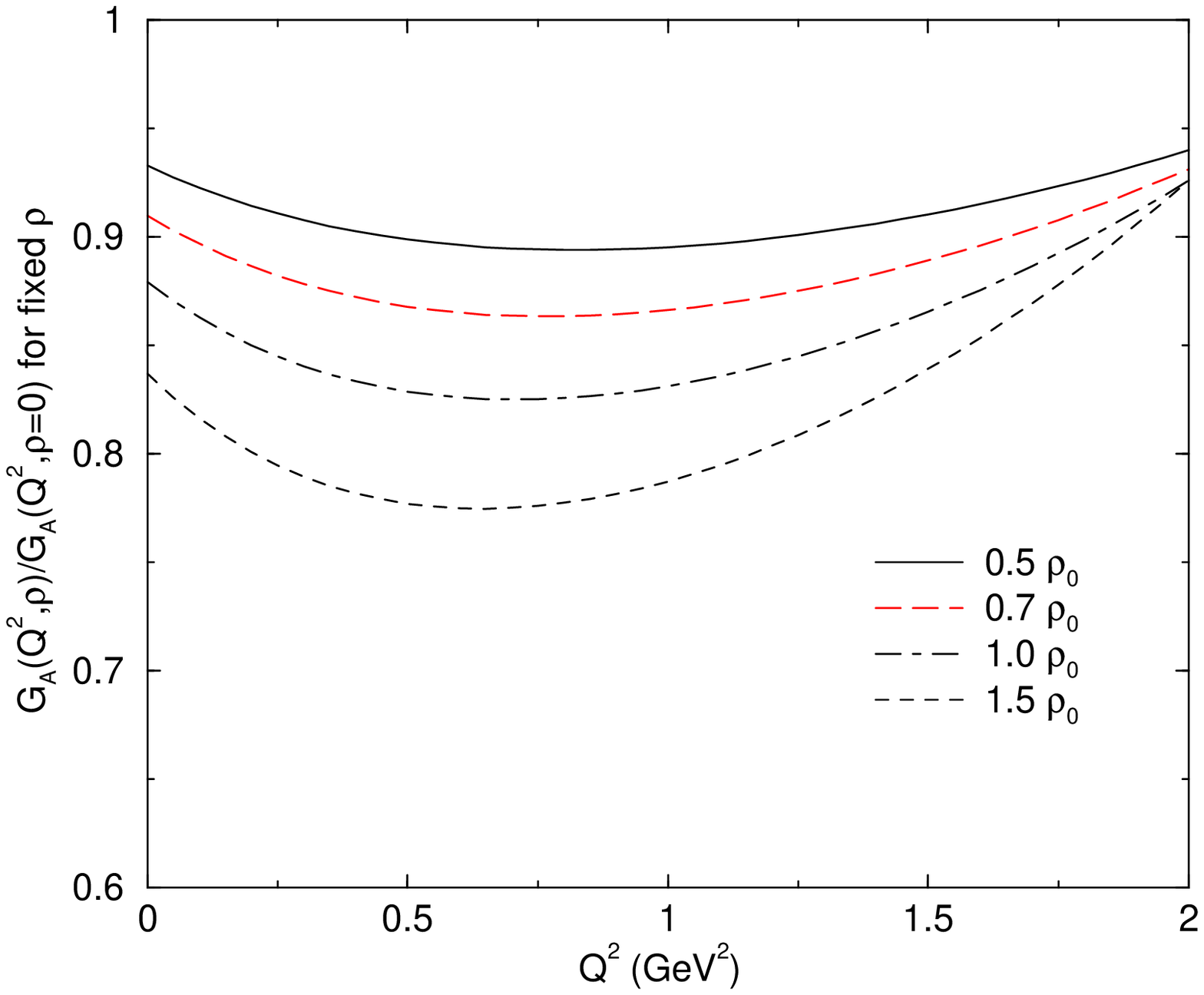,height=6.5cm,width=8.0cm}
\vspace{1em}
\caption{Free space (normalized) axial form factor $G_A(Q^2)$
calculated in the ICBM (left panel)~\protect\cite{GA}
together with experimental
data~\protect\cite{Kitagaki} summarized by a dipole form:
$G_A(Q^2)=g_A/(1+Q^2/m_A^2)^2$ with $m_A=(1.03\pm 0.04)$ GeV, 
and the ratio of in-medium to free axial form factors~\protect\cite{GA}
(right panel), where $g_A = 1.14$ is used in the ICBM calculation.
\label{axial}
}
\end{center}
\end{figure}

Similarly to the case of electromagnetic form factors, in the preferred
Breit frame the resulting bound nucleon axial form factor is
given by~\cite{GA}:
\begin{eqnarray}
G_A^*(Q^2) &=& \eta^2 G^{\rm sph\, *}_A(\eta^2 Q^2), \label{GAmedium}
\\
G_A^{sph\, *}(Q^2) &=& {5\over 3}
\int\! d^3r 
\{ 
\left[j_0^2(x_q r/R_N^*)-\beta_q^{*2}j_1^2(x_q r/R_N^*)\right]j_0(Qr) 
\nn\\
& & \hspace{20ex} 
+2\beta_q^{*2}j_1^2(x_q r/R_N^*)[j_1(Qr)/Qr] \} 
\,\,K(r)/D. 
\label{PTM}
\end{eqnarray}
In Fig.~\ref{axial} we show the (normalized) free space axial form factor
$G_A(Q^2)$ calculated in the ICBM~\cite{GA} together with the 
experimental data (left panel), and (the space component of) that
calculated at nuclear densities $\rho = (0.5,0.7,1.0,1.5) \rho_0$ with 
$\rho_0 = 0.15$~fm$^{-3}$.
%%%%%%%%%%%%%%%%%%%%%%%%%%%%%%%%%%%%%%%
%%%%%%%%%%%%%%%%%%%%%%%%%%%%%%%%%%%%%%%
%
At $Q^2 = 0$ the space component $G_A^*(Q^2=0) \equiv g_A^*$ is
quenched~\cite{gaspace} by about 10 \% at normal nuclear matter density. 
The modification calculated here may correspond to the 
``model independent part" in meson exchange language, 
where the axial current attaches itself to one of the two nucleon legs, 
but not to the exchanged meson~\cite{gaspace}. This is because  
the axial current operator in Eq.~(\ref{Acurrent}) 
is a one-body operator which operates on  
the quarks and pions belonging to a bound nucleon. 
The medium modification of the bound nucleon axial form factor
$G_A^*(Q^2)$ may be observed for instance in neutrino-nucleus scattering,
similar to that observed in the ``EMC-type'' experiments, or in a similar
experiment to the polarization transfer measurements performed on 
$^4$He~\cite{HE4,Strauch,O16}.
% (if such kinds of experiment is possible).
% ... WOULDN'T ONE NEED A PARITY-VIOLATING REACTION?
%
However, at present the experimental uncertainties seem to be too large 
to detect such medium effect directly.
We should also note that the medium modification of the parity-violating
$F_3$ structure functions of a bound nucleon in deep-inelastic 
neutrino induced reactions can be extracted using the calculated in-medium 
axial form factor $G_A^*(Q^2)$ and quark-hadron duality,
as we discuss in the next Section for the case of electromagnetic form
factors and the $F_2$ structure function.

%%%%%%%%%%%%%%%%%%%%%%%%%%%%%%%%%%%%%%%%%%%%%%%%%%%%%%%%%%%%%%%%%%%%%%%%%
\section{Quark-Hadron Duality and Nucleon Structure Functions in Medium}

A new and interesting avenue for exploring medium modifications of
hadronic observables such as form factors and structure functions is
provided by quark-hadron duality.
The relationship between form factors and structure functions, or more
generally between inclusive and exclusive processes, has been studied
in a number of contexts over the years.
Drell \& Yan \cite{DY} and West \cite{WEST} pointed out long time ago that,
simply on the basis of scaling arguments, the asymptotic behavior of
elastic electromagnetic form factors as $Q^2 \to \infty$ can be related
to the $x \to 1$ behavior of deep-inelastic structure functions.
In perturbative QCD language, this can be understood in terms of hard
gluon exchange: deep-inelastic scattering at $x \sim 1$ probes
a highly asymmetric configuration in the nucleon in which one of the
quarks goes far off-shell after the exchange of at least two hard gluons
in the initial state; elastic scattering, on the other hand, requires at
least two gluons in the final state to redistribute the large $Q^2$
absorbed by the recoiling quark \cite{LB}.

More generally, the relationship between resonance (transition) form
factors and the deep-inelastic continuum has been studied in the framework
of quark-hadron, or Bloom-Gilman, duality: the equivalence of the
averaged structure function in the resonance region and the scaling
function which describes high $W$ data.
The recent high precision Jefferson Lab data \cite{JLABF2} on the $F_2$
structure function suggests that the resonance--scaling duality also
exists locally, for each of the low-lying resonances, including
surprisingly the elastic \cite{JLABPAR}, to rather low values of $Q^2$.

%In the context of QCD, Bloom-Gilman duality can be understood within
%an operator product expansion of moments of structure functions
%\cite{RUJ,JI}: the weak $Q^2$ dependence of the low $F_2$ moments can
%be interpreted as indicating that higher twist ($1/Q^2$ suppressed)
%contributions are either small or cancel.
%However, while allowing the duality violations to be identified and
%classified according to operators of a certain twist, it does not explain
%why some higher twist matrix elements are intrinsically small.

A number of recent studies have attempted to identify the dynamical origin
of Bloom-Gilman duality using simple models of QCD \cite{DOM,IJMV,MODELS}.
It was shown, for instance, that in a harmonic oscillator basis one can
explicitly construct a smooth, scaling structure function from a set of
infinitely narrow resonances \cite{DOM,IJMV}.
%Although individual resonance contributions are suppressed by powers of
%$1/Q^2$, the number of states accessible increases with $Q^2$ so as to
%compensate the fall off, and as $Q^2 \to \infty$ quark-hadron duality
%arises from the summation over a complete set of hadronic states.
%At lower $Q^2$, however, the appearance of duality could in some cases be
%accidental, for example, because of a fortuitous cancellation of
%off-diagonal terms in the valence quark charges in the proton
%\cite{GOTT,CI,BROD}, allowing a coherent process (exclusive form factors)
%to be expressed in terms of incoherent scattering (structure functions).
Whatever the ultimate microscopic origin of Bloom-Gilman duality, for our
purposes it is sufficient to note the {\em empirical fact} that local
duality is realized in lepton-proton scattering down to
$Q^2 \sim 0.5$~GeV$^2$ at the 10-20\% level for the lowest moments of the
$F_2$ structure function.
In other words, here we are not concerned about why duality works,
but rather to assess the phenomenological consequences of the fact that
it does work.

\begin{figure}[htb]
\begin{center}
%\hspace*{-1cm}
\epsfig{file=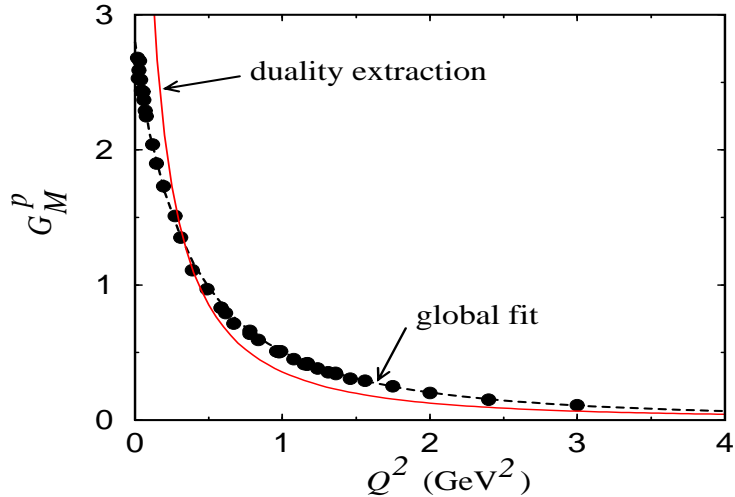,height=6.5cm,width=10cm}
\vspace{1em}
\caption{Duality extraction of the free space proton
magnetic form factor~\protect\cite{HUGS} using the quark-hadron
duality relation and the $F_2(\xi)$ parametrization in
Ref.~\protect\cite{F2Jlab}.
\label{GMdual}
}
\end{center}
\end{figure}

Motivated by the experimental verification of local duality, one can use
measured structure functions in the resonance region to directly extract
elastic form factors~\cite{RUJ}.
Conversely, empirical electromagnetic form factors at large $Q^2$ can
be used to predict the $x \to 1$ behavior of deep-inelastic structure
functions~\cite{BG,ELDUAL,QNP}.
The assumption of local duality for the elastic case implies that the area
under the elastic peak at a given $Q^2$ is equivalent to the area under
the scaling function, at much larger $Q^2$, when integrated from the pion
threshold to the elastic point \cite{BG}.
Using the local duality hypothesis, de R\'ujula et al.~\cite{RUJ}, and
more recently Ent et al.~\cite{JLABPAR}, extracted the proton's magnetic
form factor from resonance data on the $F_2$ structure function at large
$x$, finding agreement to better than 30\% over a large range of $Q^2$
($0.5 \alt Q^2 \alt 5$~GeV$^2$).
In the region $Q^2 \sim 1$--2~GeV$^2$ the agreement was at the $\sim 10\%$
level. In Fig.~\ref{GMdual} we show the extracted proton magnetic form
factor $G^p_M$ using the quark-hadron local duality 
relation~\cite{HUGS} and the $F_2^p(\xi)$ parametrization of 
Ref.~\cite{F2Jlab}, in order to estimate how reliable the quark-hadron
duality assumption may be.
%%%%%%%%%%%%%%%%%%%%%%%%%%%%%%%%%%%%%%%
%%%%%%%%%%%%%%%%%%%%%%%%%%%%%%%%%%%%%%%
%An alternative parameterization of $F_2$ was suggested in
%Ref.\cite{SIMULA}, which because of a different behavior in the unmeasured
%region $\xi \agt 0.86$, where $\xi = 2 x / (1 + \sqrt{1 + x^2/\tau})$ is
%the Nachtmann variable, with $\tau = Q^2/4M^2$, led to larger differences
%at $Q^2 \agt 4$~GeV$^2$.
%However, at $Q^2 \sim 1$~GeV$^2$ the agreement with the form factor data
%was even better here.
%As pointed out in Ref.\cite{REPLY}, data at larger $\xi$ are needed to
%constrain the structure function parameterization, and reliably extract
%the form factor at larger $Q^2$.

Applying the argument in reverse, one can formally differentiate the local
elastic duality relation \cite{BG} with respect to $Q^2$ to express the
scaling functions, evaluated at threshold,
$x = x_{\rm th} = Q^2 / (W^2_{\rm th} - M_N^2 + Q^2)$, with
$W_{\rm th} = M_N + m_\pi$, in terms of $Q^2$ derivatives of elastic form
factors.
In Refs.\cite{BG,ELDUAL} the $x \to 1$ behavior of the neutron to proton
structure function ratio was extracted from data on the elastic
electromagnetic form factors.
%(Nucleon structure functions in the $x \sim 1$ region are important as
%they reflect mechanisms for the breaking of spin-flavor SU(6) symmetry in
%the nucleon \cite{MT}.)
Extending this to the case of bound nucleons, one finds that as
$Q^2 \to \infty$ the ratio of bound to free proton structure functions is:
\begin{eqnarray}
\label{SFdual}
{ F_2^{p\, *} \over F_2^p }
&\to& { dG_M^{p\, *\, 2} / dQ^2 \over
        dG_M^{p\, 2 } / dQ^2 }\ .
\end{eqnarray}
At finite $Q^2$ there are corrections to Eq.~(\ref{SFdual}) arising from
$G_E^p$ and its derivatives, as discussed in Ref.\cite{ELDUAL}.
(In this analysis we use the full, $Q^2$ dependent expressions
\cite{ELDUAL,QNP}.)
Note that in the nuclear medium, the value of $x$ at which the pion
threshold arises is shifted:
\begin{eqnarray}
x_{\rm th} &\to& x^*_{\rm th}\
=\ \left( { m_\pi ( 2 M_N + m_\pi ) + Q^2 \over
            m_\pi (2 (M^*_N + V_N) + m_\pi ) + Q^2 }
   \right) x_{\rm th}\ ,
\end{eqnarray}
where $V_N = 3 g^q_\omega\ \bar{\omega}$ is the vector potential felt by
the nucleon and (consistent with chiral expectations and phenomenological
constraints) we have set $m_\pi^* = m_\pi$.
(See also Eq.~(\ref{omega_bar}).)
However, the difference between $x_{\rm th}$ and $x^*_{\rm th}$ has a
negligible effect on the results for most values of $x$ considered.

\begin{figure}[htb]
\begin{center}
%\hspace*{-1cm}
\epsfig{file=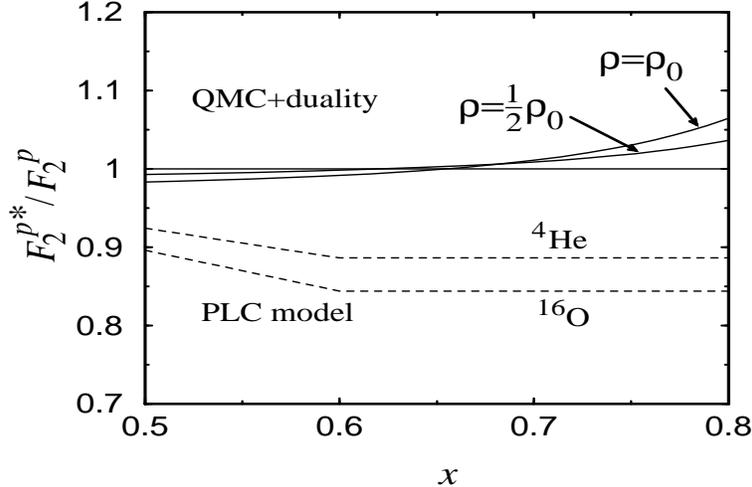,height=6.5cm,width=10cm}
\vspace{1em}
\caption{The ratios of the bound to free proton structure functions 
$F^{p*}_2/F^{p}_2$, calculated in the QMC and PLC models.
\label{F2pratio}
}
\end{center}
\end{figure}

Using the duality relations between electromagnetic form factors and
structure functions, in Fig.~\ref{F2pratio}
we plot the ratio $F_2^{p *}/F_2^p$ as
a function of $x$, with $x$ evaluated at threshold, $x = x_{\rm th}$
(solid lines).
%%%%%%%%%%%%%%%%%%%%%%%%%%%%%%%%%%%%%%%
%%%%%%%%%%%%%%%%%%%%%%%%%%%%%%%%%%%%%%%

We emphasize that since we are interested in {\em ratios} of form factors
and structure functions only, what is more relevant for our analysis is
not the degree to which local duality holds for the {\em absolute}
structure functions, but rather the {\em relative} change in the duality
approximation between free and bound protons.
Note that at threshold the range of $Q^2$ spanned between $x=0.5$ and
$x=0.8$ is $Q^2 \approx 0.3$--1.1~GeV$^2$.
Over the range $0.5 \alt x \alt 0.75$ the effect is almost negligible,
with the deviation of the ratio from unity being $\alt 1\%$ for
$\rho={1\over 2}\rho_0$ and $\alt 2\%$ for $\rho=\rho_0$.
For $x \agt 0.8$ the effect increases to $\sim 5\%$, although, since
larger $x$ corresponds to larger $Q^2$, the analysis in terms of the QMC
model is less reliable here.
However, in the region where the analysis can be considered reliable, the
results based on the bound nucleon form factors inferred from the
polarization transfer data~\cite{HE4,Strauch} and local duality imply that the
nucleon structure function undergoes very little modification in medium.

It is instructive to contrast this result with models of the EMC effect
in which there is a large medium modification of nucleon structure.
For example, let us consider the model of Ref.~\cite{FS}, where it is
assumed that for large $x$ the dominant contribution to the structure
function is given by the point-like configurations (PLC) of partons which
interact weakly with the other nucleons.
The suppression of this component in a bound nucleon is assumed to be the
main source of the EMC effect.
This model represents one of the extreme possibilities that the EMC
effect is solely the result of deformation of the wave function of bound
nucleons, without attributing any contribution to nuclear pions or other
effects associated with nuclear binding \cite{MSS}.
%Given that this model has been so successfully applied to describe the
%nuclear EMC effect, it is clearly important to examine its consequences
%elsewhere.

The deformation of the bound nucleon structure function in the PLC
suppression model is governed by the function \cite{FS}:
\begin{eqnarray}
\label{delta}
\delta(k) &=& 1 - 2 (k^2/2M + \epsilon_A)/\Delta E_A\ ,
\end{eqnarray}
where $k$ is the bound nucleon momentum, $\epsilon_A$ is the nuclear
binding energy, and $\Delta E_A \sim 0.3$--0.6~GeV is a nucleon
excitation energy in the nucleus.
For $x \agt 0.6$ the ratio of bound to free nucleon structure functions
is then given by \cite{FS}:
\begin{eqnarray}
\label{plc}
{ F_2^{N\, *}(k, x) \over F_2^N(x) }
&=& \delta(k)\ .
\end{eqnarray}
The $x$ dependence of the suppression effect is based on the assumption
that the point-like configuration contribution in the nucleon wave
function is negligible at $x \alt 0.3$ ($F_2^{N\, *}/F_2^N = 1$), and for
$0.3 \alt x \alt 0.6$ one linearly interpolates between these values
\cite{FS}.
The results for $^4$He and $^{16}$O are shown 
in Fig.~\ref{F2pratio} (dashed lines)
for the average values of nucleon momentum, $\langle k^2 \rangle$, in
each nucleus.
The effect is a suppression of order 20\% in the ratio
$F_2^{N\, *}/F_2^N$ for $x \sim 0.6$--0.7.
In contrast, the ratios extracted on the basis of duality, using the QMC
model constrained by the $^4$He polarization transfer data~\cite{HE4,Strauch},
show almost no suppression ($\alt 1$--2\%) in this region.
Thus, for $^4$He, the effect in the PLC suppression model is an order
of magnitude too large at $x \sim 0.6$, and has the opposite sign for
$x \agt 0.65$.

Although the results extracted from the polarization transfer
measurements~\cite{HE4,Strauch} rely on the assumption of local duality, we
stress that the corrections to duality have been found to be typically
less than 20\% for $0.5 \alt Q^2 \alt 2$~GeV$^2$ \cite{JLABF2,SIMULA}.
The results therefore appear to rule out large bound structure function
modifications, such as those assumed in the point-like configuration
suppression model \cite{FS}, and instead point to a small medium
modification of the intrinsic nucleon structure, which is complemented
by standard many-body nuclear effects.

\begin{figure}[htb]
\begin{center}
%\hspace*{-1cm}
\epsfig{file=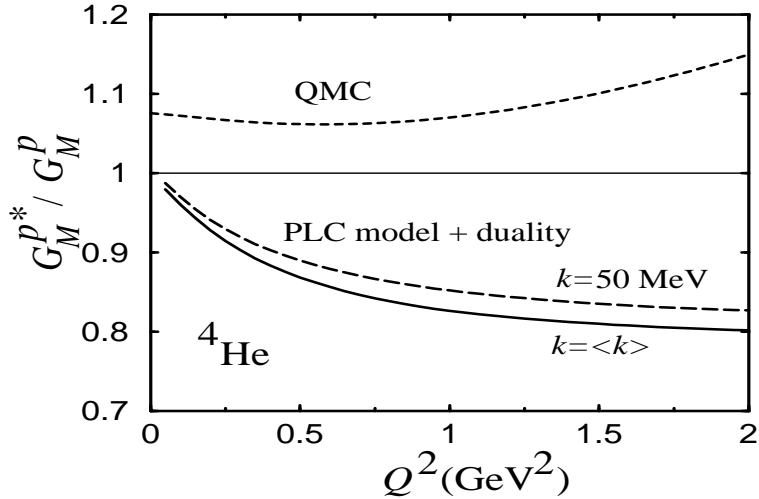,height=6.5cm,width=10.0cm}
\vspace{1em}
\caption{The ratios of the bound proton to free magnetic form factors
$G^{p*}_M/G^{p}_M$, calculated in the QMC and PLC models.
\label{GMpratio}
}
\end{center}
\end{figure}

Nevertheless, given the large differences between the theoretical
predictions, data on nuclear structure functions at large $x$ would be
extremely valuable in discriminating between these scenarios.
Possible insights into the medium modifications of bound nucleon
structure functions may be garnered from semi-inclusive deep-inelastic
scattering experiments from deuteron, by studying the spectrum of
tagged spectator nucleons \cite{E94102}.
As a consistency check on the analysis, one can also examine the change
in the form factor of a bound nucleon that would be implied by the
corresponding change in the structure function in medium.
Namely, from the local duality relation \cite{RUJ,QNP}:
\begin{eqnarray}
\label{elint}
\left[ G_M^p(Q^2) \right]^2
&\approx& { 2 - \xi_0 \over \xi_0^2 }
          { (1 + \tau) \over (1/\mu_p^2 + \tau) }
          \int_{\xi_{\rm th}}^1 d\xi\ F_2^p(\xi)\ ,
\end{eqnarray}
one can extract the magnetic form factor by integrating the $F_2^p(\xi)$
structure function over $\xi$
between threshold, $\xi = \xi_{\rm th}$, and $\xi=1$.
Here $\xi_0 = \xi(x=1)$, $\mu_p$ is the proton magnetic moment, 
and $\xi = 2x/(1+\sqrt{1+x^2/\tau})$ with 
$\tau = Q^2/4M_N^2$ and $x$ the Bjorken variable.
In Fig.~\ref{GMpratio} we show the PLC model predictions 
for the ratio of the magnetic
form factor of a proton bound in $^4$He to that in vacuum, derived from
Eqs.~(\ref{plc}) and (\ref{elint}), using the parameterization for
$F_2^p(\xi)$ from Ref.\cite{JLABPAR}, and an estimate for the in-medium
value of $\mu_p^*$ from Ref.\cite{escatt,He3,QMCB,HeO}.
Taking the average nucleon momentum in the $^4$He nucleus,
$k = \langle k \rangle \approx 135$~MeV, %% K.E. = 9.8 MeV
the result is a suppression of about 20\% in
the ratio $G^{p *}_M/G^p_M$ at $Q^2 \sim 1$--2~GeV$^2$ (solid curve).
Since the structure function suppression in the PLC model depends on the
nucleon momentum (Eq.~(\ref{delta})), we also show the resulting form
factor ratio for a momentum typical in the $(\vec e, e' \vec p)$
experiment, $k = 50$~MeV (long dashed).
As expected, the effect is reduced, however, it is still of the order
15\% since the suppression also depends on the binding energy, as well
as on the nucleon mass, which changes with density rather than with momentum.
In contrast, the QMC calculation, which is consistent with the MAMI
$^4$He quasi-elastic data, and Jlab polarization transfer measurements 
on $^4$He~\cite{HE4,Strauch}, produces a ratio which is typically 5--10\%
{\it larger} than unity (dashed).
Without a very large compensating change in the in-medium electric form
factor of the proton (which seems to be excluded by $y$-scaling
constraints), the behavior of the magnetic form factor implied by the
``PLC model $+$ duality" would produce a large {\em enhancement} of the
polarization transfer ratio, rather than the observed small 
suppression~\cite{HE4,Strauch} (see Eq.~(\ref{GEpGMp}) and
Fig.~\ref{JlabQMC}).

%These results have other important practical ramifications.
%For instance, the PLC suppression model was used recently \cite{SSS} to
%argue that the EMC effects in $^3$He and $^3$H differ significantly at
%large $x$, in contrast to calculations \cite{AFNAN,PACE} based on
%conventional nuclear physics using well-established bound state wave
%functions which show only small differences.
%Based on the findings presented here, one would conclude that the
%conventional nuclear physics description of the $^3$He/$^3$H system
%should indeed be a reliable starting point for nuclear structure function
%calculations, as the available evidence suggests little room for large
%off-shell corrections.
%%%%%%%%%%%%%%%%%%%%%%%%%%%%%%%%%%%%%%%
%%%%%%%%%%%%%%%%%%%%%%%%%%%%%%%%%%%%%%%

%%%%%%%%%%%%%%%%%%%%%%%%%%%%%%%%%%%%%%%%%%%%%%%%%%%%%%%%%%%%%%%%%%%%%%%%
\section{Summary}

In this article we have discussed the medium modification of the internal
structure of bound nucleons due to the change of quark response to 
the nuclear environment. Recent experimental
results~\cite{HE4,Strauch} from polarized proton
knockout reactions off $^4$He nuclei have provided the first direct
evidence for a possible small but nonzero modification of the proton
electromagnetic form factors in the nuclear medium.
The analyses in Refs.~\cite{HE4,Strauch} 
found that when compared with conventional
nuclear calculations, the medium modifications observed in the $^4$He
data could only be accounted for within models in which the conventional
descriptions are supplemented by a small modification of the nucleon form
factors in medium calculated in the QMC
model~\cite{escatt,He3,QMCB,HeO,Guichon,Guichonf,Saitof}
(see also Ref.~\cite{CELENZA}).

We have also analyzed the medium modification of the bound nucleon 
axial form factor $G_A(Q^2)$.
%
% and have discussed an application 
% in extracting the medium modification of bound nucleon 
% structure function $F_3$, using quark-hadron duality.
%
The modification of the axial form factor may be observed for instance
in future high precision neutrino--nucleus scattering experiments
\cite{NUMI}.

Finally, we have examined the consequences of quark-hadron duality
applied to nucleons in the nuclear medium.
Utilizing the experimental results~\cite{HE4,Strauch} 
we have used local duality to relate {\em model-independently} the medium
modified electromagnetic form factors to the change in 
the intrinsic structure function of a bound
proton.
While the results rely on the validity of quark-hadron duality, the
empirical evidence suggests that for low moments of the proton's $F_2$
structure function the duality violations due to higher twist corrections
are $\alt 20\%$ for $Q^2 \agt 0.5$~GeV$^2$ \cite{JLABF2}, and decrease
with increasing $Q^2$.
In the context of the QMC model, the change in nucleon form factors
allowed by the data imply a modification of the in-medium structure
function of $\alt 1$--2\% at $0.5 \alt x \alt 0.75$ for all nuclear
densities between nuclear matter densities, $\rho=\rho_0$, and
$\rho={1\over 2}\rho_0$.

The results place rather strong constraints on models of the nuclear EMC
effect, especially on models which assume that the EMC effect arises from
a large deformation of the nucleon structure in medium.
%For example, we find that the PLC suppression model \cite{FS} predicts
%an effect which is about an order of magnitude larger than that allowed
%by the data~\cite{HE4,Strauch}, and has a different sign.
%
While suggesting the need for explicit quark degrees of freedom in the
nucleus, our findings appear to disfavor models with large medium
modifications of structure functions as viable explanations for the
nuclear EMC effect, although it would be desirable to have more data
on a variety of nuclei and in different kinematical regions.
%The recently completed Jefferson Lab $^4$He polarization transfer
%experiment, which covered a large range of $Q^2$, between 0.5~GeV$^2$
%and 2.6~GeV$^2$ \cite{E93049}, should provide valuable additional
%information.
%Final results \cite{Strauch} indicate that the lowest $Q^2$ point
%is in very good agreement with the Mainz $Q^2=0.4$~GeV$^2$ data point,
%which provides further support for the QMC description.
A such proposed experiment \cite{P01013} at Jefferson Lab on $^{16}$O
at $Q^2=0.8$~GeV$^2$, which would make use of other, high precision cross
section data at this momentum transfer, would have about 15 times the
statistics of the original commissioning experiment~\cite{O16}.
This would enable a more thorough comparison of the medium dependence
of form factors and structure functions for different nuclei.

%These results have other important practical ramifications.
%For instance, the PLC suppression model was used recently \cite{SSS} to
%argue that the EMC effects in $^3$He and $^3$H differ significantly at
%large $x$, in contrast to calculations \cite{AFNAN,PACE} based on
%conventional nuclear physics using well-established bound state wave
%functions which show only small differences.
%Based on the findings presented here, one would conclude that the
%conventional nuclear physics description of the $^3$He/$^3$H system
%should indeed be a reliable starting point for nuclear structure function
%calculations, as the available evidence suggests little room for large
%off-shell corrections.
%Finally, let us stress that quark-hadron duality is a powerful tool with
%which to simultaneously study the medium dependence of both exclusive
%and inclusive observables, and thus provides an extremely valuable guide
%towards a consistent picture of the effects of the nuclear environment on
%nucleon substructure.

\vspace{1em}
%%%%%%%%%%%%%%%%%%%%%%%%%%%%%%%%%%%%%%%
\noindent
{\bf Acknowledgment: }\\
%%%%%%%%%%%%%%%%%%%%%%%%%%%%%%%%%%%%%%%
% We would like to acknowledge a warm hospitality at
% CSSM, University of Adelaide, Australia. 
KT is supported by the Forschungszentrum-J\"ulich, 
contract No. 41445282 (COSY-058).
DHL is grateful to the Y.C. Tang Disciplinary Development 
Fund in Zhejiang University. 
His work was supported in part by the National Natural Science Fund 
of China and by the Australian Research Council.
WM is supported by the U.S. 
Department of Energy contract~\mbox{DE-AC05-84ER40150}, under which the
Southeastern Universities Research Association (SURA) operates the
Thomas Jefferson National Accelerator Facility (Jefferson Lab).

%%%%%%%%%%%%%%%%%%%%%%%%%%%%%%%%%%%%%%%
%%%%%% References
%\begin{thebibliography}{99}
%%%%%%%%%%%%%%%%%%%%%%%%%%%%%%%

%\bibitem{SSS}
%M.M.~Sargsian, S.~Simula and M.I.~Strikman,
%Phys. Rev. C {\bf 66}, 024001 (2002).
%
%\bibitem{AFNAN}
%I.R.~Afnan, F.~Bissey, J.~Gomez, A.T.~Katramatou, W.~Melnitchouk,
%G.G.~Petratos and A.W.~Thomas,
%Phys. Lett. B {\bf 493}, 36 (2000).
%
%\bibitem{PACE}
%E.~Pace, G.~Salme, S.~Scopetta and A.~Kievsky,
%Phys. Rev. C {\bf 64}, 055203 (2001).
%
%%%%%%%%%%%%%%%%%%%%%%%%%%%%%%%
%\end{thebibliography}
%
%%%%%%%%%%%%%%%%%%%%%%%%%%%%%%%
\end{document}